\documentclass[a4paper]{llncs}
\usepackage[T1]{fontenc}
\usepackage[center]{caption2}
\usepackage{times}
\usepackage{epsfig}
\usepackage{graphicx}

\usepackage{subfigure}

\begin{document}

\title{FIPA-based Interoperable Agent Mobility Proposal}

\author{Jordi Cucurull, Ramon Mart\'\i, Sergi Robles,
        Joan Borrell, Guillermo Navarro}

\institute{Department of Information and Communications Engineering,\\
 Universitat Aut\`onoma de Barcelona,\\
  08193 Bellaterra - Spain \\
  \email{\{jcucurull, rmarti, srobles, jborrell, gnavarro\}@deic.uab.cat}
}

\maketitle

\begin{abstract}
  This paper presents a proposal for a flexible agent mobility
  architecture based on IEEE-FIPA standards and intended to be one of
  them. This proposal is a first step towards interoperable mobility
  mechanisms, which are needed for future agent migration between
  different kinds of platforms. Our proposal is presented as a
  flexible and robust architecture that has been successfully
  implemented in the JADE and AgentScape platforms. It is based on an
  open set of protocols, allowing new protocols and future
  improvements to be accommodated in the architecture. With this
  proposal we demonstrate that a standard architecture for agent
  mobility capable of supporting several agent platforms can be
  defined and implemented.
\end{abstract}


\section{Introduction}
\label{sec:introduction}

Mobile agents are software agents with the ability of travelling from
one execution environment (platform) to another across a computer
network~\cite{white96}. Their introduction and studying during the
last decade have opened an interesting research field with new
applications~\cite{VRCCNM525} and paradigms. Unfortunately, mobile
agents have also raised some issues regarding
security~\cite{JK00,roth01} and
interoperability~\cite{Pinsdorf:2002a}, which are still unsolved.

The basic operations behind an agent migration are suspending the
execution of the agent in its current execution platform, transferring
the agent code, data and state to another platform, and resuming the
execution of the agent in the same state it was before the migration
took place. The actual migration is far more complex, since there are
many issues that must be taken into account, ranging from security and
access control mechanisms to interoperability of different hardware,
languages, and agent platforms.

Several contributions in the agent mobility field have been made
comprising different agent platforms such as Aglets~\cite{lo98},
D'Agents~\cite{586291}, or Ara~\cite{732412} just to mention some.
Despite the number of proposed and developed mobile agent platforms,
they happen to be incompatible between them. In an attempt to solve,
or at least to minimise, the problem of incompatibility and
interoperability some organisations have driven the development of
standards. The first effort was the OMG-MASIF~\cite{MASIF}
specification written by the OMG group, an attempt to standardise
mobile agent APIs. The second, and most popular nowadays, was taken by
the IEEE-FIPA organisation, which created several specifications
related to agent communications~\cite{FIPA023,FIPA061}, including even
one for agent mobility~\cite{FIPA087} that was deprecated due to a
lack of implementations.

Besides standardisation, several interesting works have recently
appeared in the field of agent mobility. One of this works is the
Kalong architecture~\cite{br05}, a mobility module used in the Tracy
platform which is exportable to other platforms~\cite{btk05}. It is a
complete migration system on its own, using different techniques to
optimise the agent code transfer. Nevertheless it is not based on any
known standard. Another interesting work is the AgentScape
Operating System~\cite{Overeinder.ea:04}, which is a mobile agent
platform focused in scalability and security. It has a separate
subsystem which can register and move agent code and data, making them
usable for other platforms. However, AgentScape does not pursue the
standardisation of agent facilities. And finally, there is our
previous contribution to the field of mobility, which is the
implementation of a mobility service for the JADE platform \cite{JADE}
following IEEE-FIPA specifications~\cite{acmnr0906}. This one is
currently being used by the whole community of JADE users, and it is
the basis for the work described in this paper.  This mobility service
is very rigid and requires a fixed set of protocols, which might not
be suitable for all situations.  This paper presents a new proposal
for the future standardisation of a flexible agent migration process
as a new IEEE-FIPA specification. The aim of this proposal is twofold.
Firstly, it sets the foundations for an interoperable migration among
different agent platforms. Secondly, it tries to be completely
flexible allowing the integration of future migration protocols.

The paper is organised as follows: Section~\ref{sec:former-migration}
shows a brief description of the former migration;
Section~\ref{sec:main-protocol} describes the new migration proposal;
and Section~\ref{sec:push-transfer-protocol} presents a simple
transfer protocol for the first implementations. Finally,
Section~\ref{sec:conclusions} summarises our conclusions and future
work.

\section{Former migration}
\label{sec:former-migration}

\begin{figure}
  \centering
    \includegraphics[width=6cm]{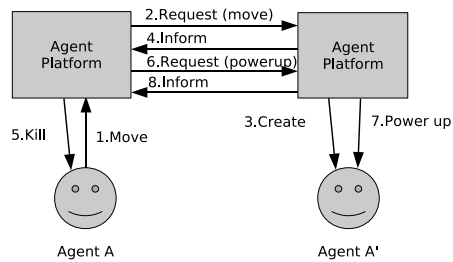}
    \caption{FIPA-based mobility protocol.}
    \label{fig:protocol}
\end{figure}

The proposal of the migration architecture presented in this paper is
the natural evolution of the work presented in~\cite{acmnr0906}, a
simple inter-platform mobility service implemented over the JADE agent
platform. That former proposal was partially based on the FIPA Agent
Management Support for Mobility Specification~\cite{FIPA087}, and
followed the FIPA-Request Interaction Protocol~\cite{FIPA026} with two
steps: a first one to transfer the agent, and a second one to start
it. Figure~\ref{fig:protocol} illustrates its sequence of messages.

The purpose of this first implementation was to experiment with the
old FIPA specification, to identify drawbacks, to evaluate the performance
of an ACL-based mobility and to evaluate its flexibility for interoperating
with different types of platforms.

\section{Multi-protocol based architecture proposal}
\label{sec:main-protocol}

Our proposal defines a new architecture to migrate agents among
platforms using some of the agent communication standards proposed by
the IEEE-FIPA organisation. This proposal aims a double objective. On
one hand, it tries to set the grounds for an interoperable migration
between different types of agent platforms. On the other hand, it
defines a flexible framework in which future migration protocols will
seamlessly integrate.

\begin{figure}[!ht]
  \begin{minipage}[t]{0.5\linewidth}
    \centering
        \includegraphics[width=4.8cm]{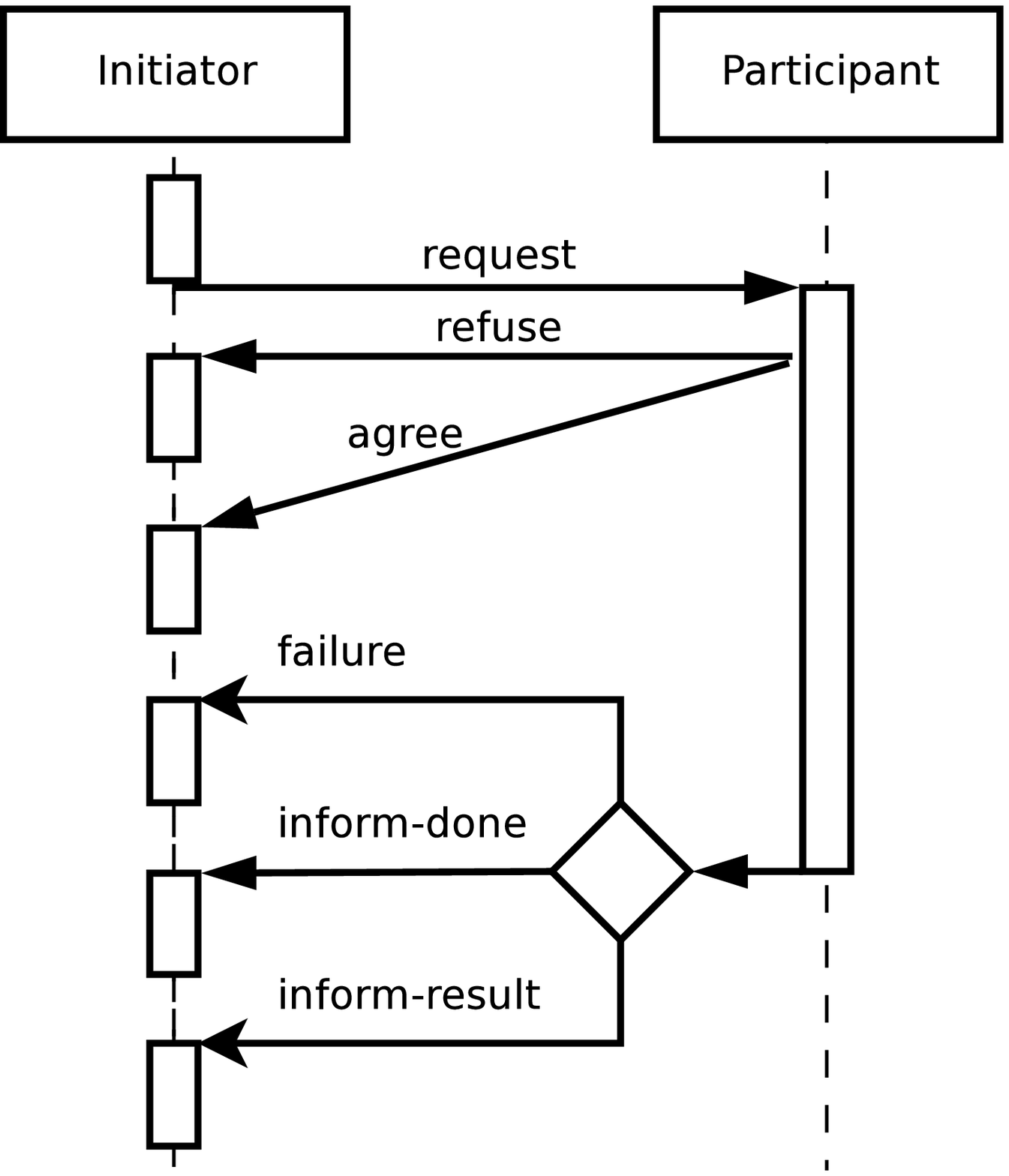}
        \caption{FIPA Request Interaction protocol.}
        \label{fig:fipa-request}
  \end{minipage}%
  \begin{minipage}[t]{0.5\linewidth}
    \centering
    \includegraphics[width=4.8cm]{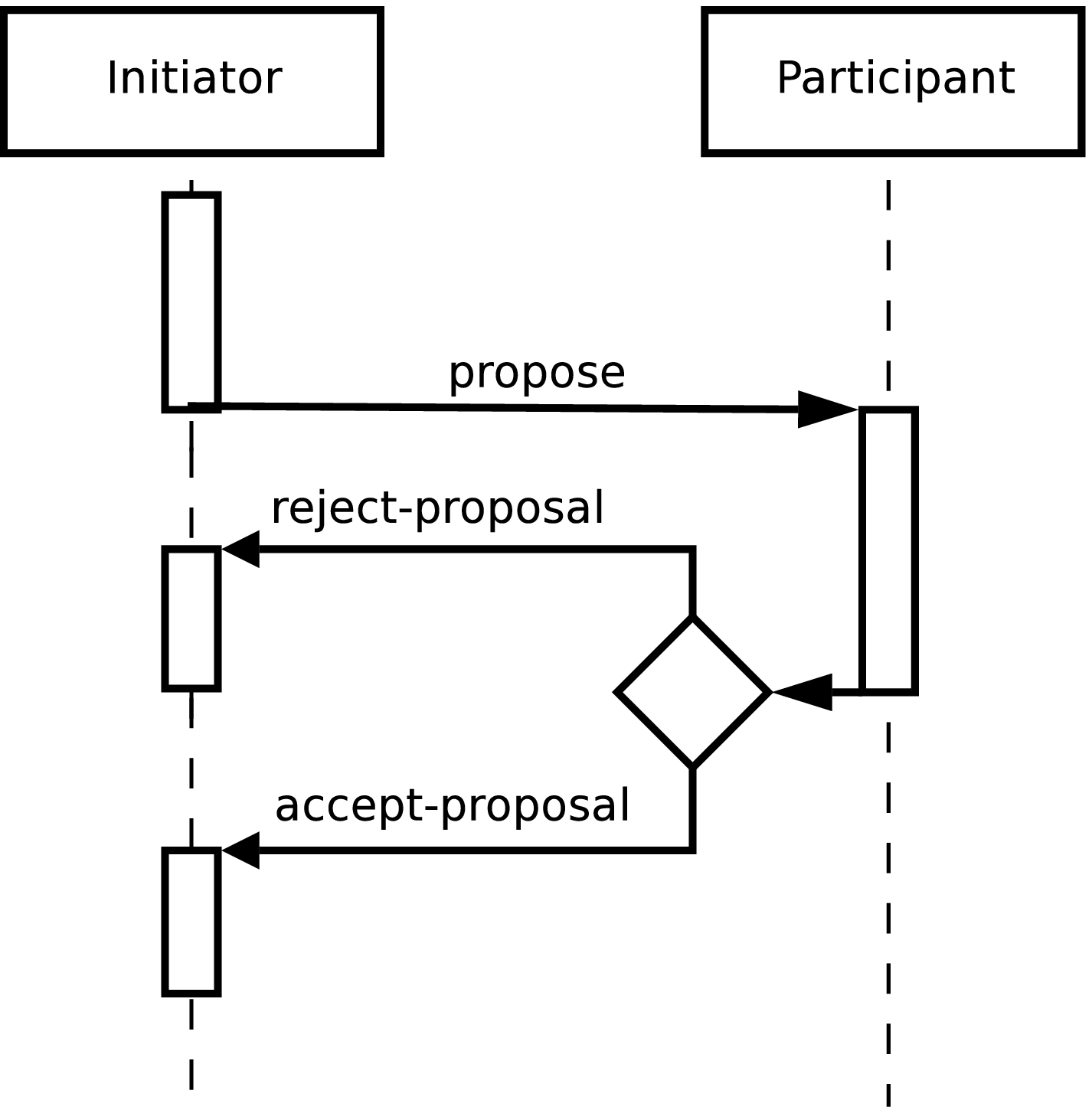}
    \caption{FIPA Proposal Interaction protocol.}
    \label{fig:fipa-proposal}
  \end{minipage}
\end{figure}

\begin{figure}[!ht]
  \centering
  \includegraphics[width=8cm]{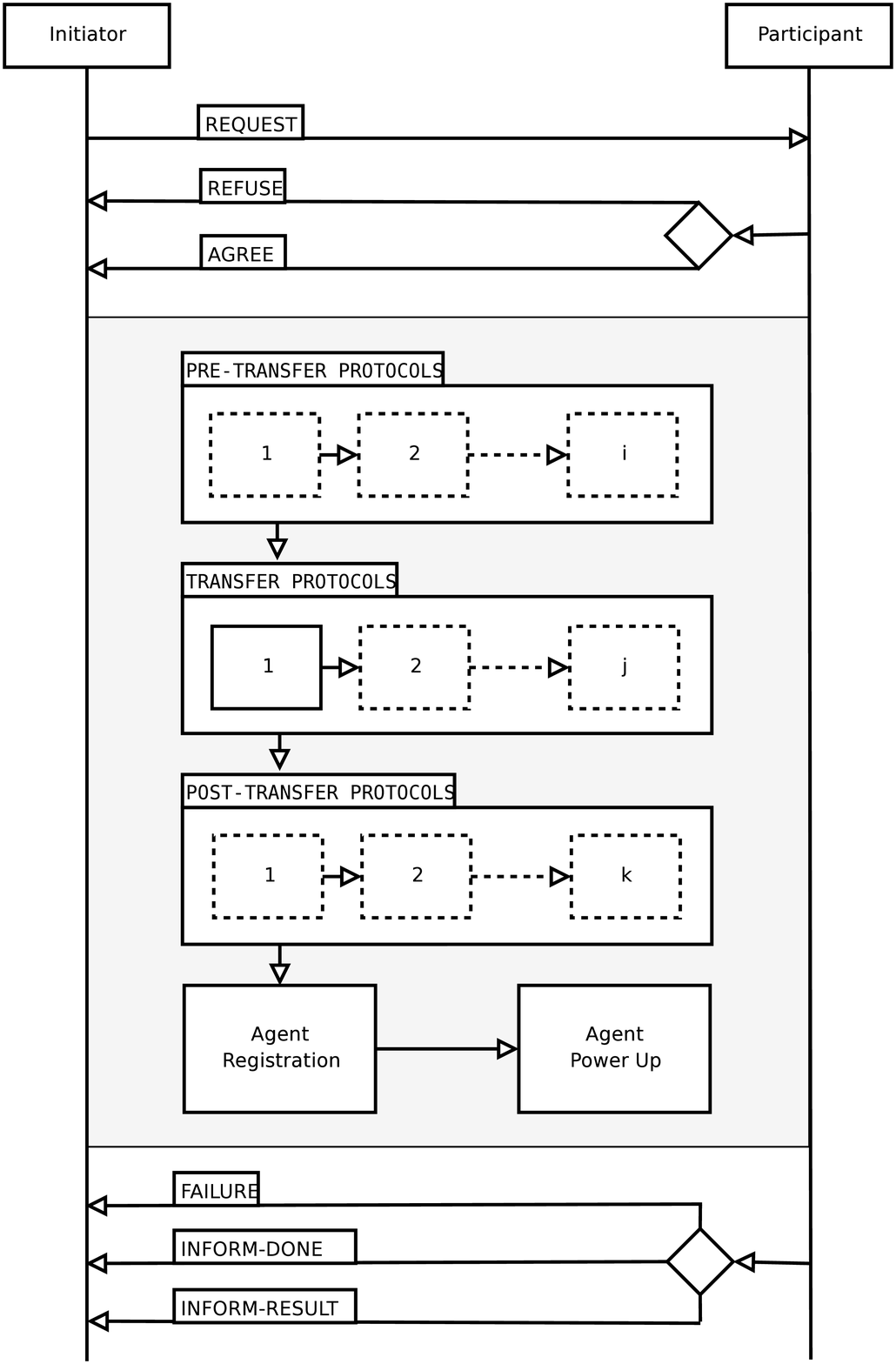}
  \caption{Whole migration process.}
  \label{fig:whole-migration}
\end{figure}

The migration process %
is split into five different steps %
contained inside a \emph{Main} protocol that follows an IEEE FIPA
Request Interaction Protocol (see Figure~\ref{fig:fipa-request}),
which form together the whole migration process (this is illustrated
in Figure~\ref{fig:whole-migration}). The same ontology is shared by
all protocols in the proposal: the \emph{migration-ontology}. All the
architecture is managed by an agent called \emph{amm} (Agent Mobility
Manager).

The steps proposed for the migration process are: Pre-Transfer
($s_{pre-transfer}$), Transfer ($s_{transfer}$), Post-Transfer
($s_{post-transfer}$), Agent Registration ($s_{registration}$) and
Agent Power Up ($s_{powerup}$). Each one of the first three steps is
implemented by an open set of protocols %
that is specified in the first sent message.  These protocols are out
of the scope of this proposal. The last two steps (registration and
powering up) are implemented by two fixed protocols, defined later,
which are mandatory in all platforms supporting this proposal. Each
migration process must execute all the steps in the listed order.

Each one of the above mentioned protocols: %
must have a well defined functionality suitable with one of the steps;
is independent from the rest; can be composed of many stages; should
use one or more ontologies and interaction protocols; and must have a
well-known name. 

Since this proposal is intended to be a new IEEE-FIPA standard, it
uses several of the current specifications of this organisation. First
of all, the mobile agent conforms to the IEEE-FIPA Agent Management
Specification \cite{FIPA023}, with the standard lifecycle and the
agent naming conventions defined there. Furthermore, all the messages
used in this proposal rely on the ACL Message specification defined in
\cite{FIPA061} and take advantage of using IEEE-FIPA Interaction
Protocols.  These protocols are used because they provide a more
consistent message exchange between the involved platforms and,
moreover, because they reflect and delimit the different parts
comprising the migration. Note that although the aim of this
architecture is to use the IEEE-FIPA specifications and its
interaction protocols, developers are free to implement their own
protocols using alternative technologies.

\subsection{Main protocol}
\label{sec:general-protocol}

The \textit{Main} protocol starts the migration process and manages
the rest of protocols. As mentioned before, the whole migration
process follows an IEEE-FIPA Request Interaction
Protocol~\cite{FIPA026} (see Figure~\ref{fig:fipa-request}).
The first message sent by the initiator requests the destination
platform to start the process. It contains a \texttt{move} or
\texttt{clone} action (see Table~\ref{tab:move-action}) together with
the \emph{Mobile Agent Description} (see Table~\ref{tab:dissmobont1}),
which contains information about the agent compatibility and the set
of protocols to be used, including the \emph{Mobile Agent Profile}
used as defined in~\cite{FIPA087}. Furthermore, a unique session ID is
generated to uniquely identify the whole migration process.

An immediate response is received from the receiver agreeing or
refusing the migration request (aborting the whole migration in this
last case). In case of agreement both participants collaborate by
running the specific migration steps they have agreed (each one using
the protocols requested in the previous \texttt{clone} or
\texttt{move} action). Finally, the result of the whole migration
process is received by an \texttt{inform} or a \texttt{failure}
message. It should be noted that a migration process is considered
failed when any protocol in any of its steps fails.

\begin{table*}
  \centering
  \begin{tabular}{|p{60pt}|p{275pt}|}
    \hline
    \textbf{Function} & move / clone \\
    \hline
    \textbf{Ontology} & migration-ontology \\
    \hline
    \textbf{Supported by} & amm \\
    \hline
    \textbf{Description} & Request to start an agent migration/cloning process. \\
    \hline
    \textbf{Domain} & mobile-agent-description \\
    \hline
  \end{tabular}
  \caption{Migration Ontology: Move / Clone action}
  \label{tab:move-action}
\end{table*}

\begin{table*}
  \centering
  \begin{tabular}{|p{60pt}|p{150pt}|p{50pt}|p{75pt}|}
    \hline
    \multicolumn{1}{|l|}{\textbf{Frame}} &
    \multicolumn{3}{|l|}{mobile-agent-description} \\
    \hline
    \multicolumn{1}{|l|}{\textbf{Ontology}} &
    \multicolumn{3}{|l|}{migration-ontology} \\
    \hline
    \textbf{Parameter}&\textbf{Description}&\textbf{Presence}&\textbf{Type} \\
    \hline
    name & The agent identifier & Mandatory & agent-identifier \\
    \hline
    agent-profile & List of agent requirements. & Mandatory & mobile-agent-profile \\
    \hline
    agent-version &  Agent version. & Optional & String \\
    \hline
    pre-transfer & Pre-transfer protocols chosen & Optional & Set of String \\
    \hline
    transfer & Transfer protocols chosen & Mandatory & Set of String \\
    \hline
    post-transfer & Post-transfer protocols chosen & Optional & Set of String \\
    \hline
  \end{tabular}
  \caption{Migration Ontology: Mobile agent description}
  \label{tab:dissmobont1}
\end{table*}

\subsection{Pre-Transfer, Transfer and Post-Transfer steps}
\label{sec:three-steps}

The flexibility of our architecture comes from the Pre-Transfer,
Transfer, and Post-Transfer steps. In each migration process, a subset
of protocols are chosen and specified by the migrating agent in the
initial \texttt{request} message. If the migration is agreed, then the
protocols are executed in the same order as specified in the first
message. The general functionality of these protocols is defined by
this proposal, but the specific protocols must be proposed and
standardised apart.  At least one transfer protocol should be provided
to do a complete migration (one is proposed in
Section~\ref{sec:push-transfer-protocol}.

The three steps revolve around the transfer of the agent, which is the
central part of a migration process. Next, a brief description of
these steps follows:

\begin{itemize}
\item \textbf{Pre-Transfer}: In this step the set of protocols needed
  before the agent transfer is run. The protocols used in this step
  can be related to authentication, authorisation, resource agreement,
  etc. For a specific migration, there can be zero or more
  pre-transfer protocols.
\item \textbf{Transfer}: The set of protocols to transfer the agent
  code, data and state is run in this step.  Different kinds of
  protocols allow to follow different migration strategies (push, on
  demand, etc.). There should be at least one transfer protocol in
  this step.
\item \textbf{Post-Transfer}: In this step the set of protocols needed
  after the agent transferring is run. The protocols in this step can
  be used for authorisation, agent data transfer, etc. This step can
  have zero or more post-transfer protocols.
\end{itemize}

\subsection{Agent Registration step}
\label{sec:agent-registration}

In the Agent Registration step ($s_{registration}$), the agent is
rebuilt and registered in the destination platform. Then, if the
registration is successful the agent should be killed on the origin
platform.

In this case only one pre-defined protocol is allowed to be used,
the Agent Registration Protocol ($p_{registration}$), identified by
the ``registration-protocol-v1'' string.

This protocol uses the simplified IEEE-FIPA Request Interaction
Protocol where the \texttt{agree} and \texttt{refuse} messages are not
used. The first message contains the agent identifier to be registered
over the \texttt{register} action (see
Table~\ref{tab:register-action}). Then, as a result, it is expected an
\texttt{inform-done} message if the operation succeeds or a
\texttt{failure} one otherwise. In the first case, the agent in the
source platform should be killed.

\begin{table*}
  \centering
  \begin{tabular}{|p{60pt}|p{275pt}|}
    \hline
    \textbf{Function} & register / power-up \\
    \hline
    \textbf{Ontology} & migration-ontology \\
    \hline
    \textbf{Supported by} & amm \\
    \hline
    \textbf{Description} & Request to register / power-up an agent in a remote platform. \\
    \hline
    \textbf{Domain} & agent-identifier \\
    \hline
  \end{tabular}
  \caption{Migration Ontology: Register / Power Up action}
  \label{tab:register-action}
\end{table*}

\subsection{Agent Power Up step}
\label{sec:agent-power-up}

In the Agent Power Up step ($s_{power-up}$), the destination platform
is requested to resume the execution of the received agent. Only one
protocol is allowed to be used here,
the Agent Power Up Protocol ($p_{power-up}$), identified by the
string ``power-up-protocol-v1''.

This protocol also uses the simplified version of the IEEE-FIPA
Request Interaction Protocol. The first message contains the
\texttt{power-up} action (see Table~\ref{tab:register-action}) and the
agent identifier, in order to confirm the agent to start. Then, an
\texttt{inform-done} is returned if the agent has been correctly
started in the destination platform, or a \texttt{failure} if an
exception has been thrown.

\section{Push Transfer Protocol}
\label{sec:push-transfer-protocol}

This section presents the Push Transfer Protocol, proposed outside the
main architecture of the migration process. This transfer protocol is
based on the simple transfer protocol presented in the former
migration mechanism explained in Section~\ref{sec:former-migration}.
It is identified by the ``push-transfer-protocol-v1'' well-known name.

It is called Push Transfer Protocol because all the agent code along
with the agent data and state (this last one only if needed), is
directly sent from the platform where the agent resides.
Furthermore, this protocol allows to save up network bandwidth because
the agent code is only sent in case that the destination platform does
not have a copy of it.

\begin{figure}[!ht]
  \centering
  \includegraphics[width=9cm]{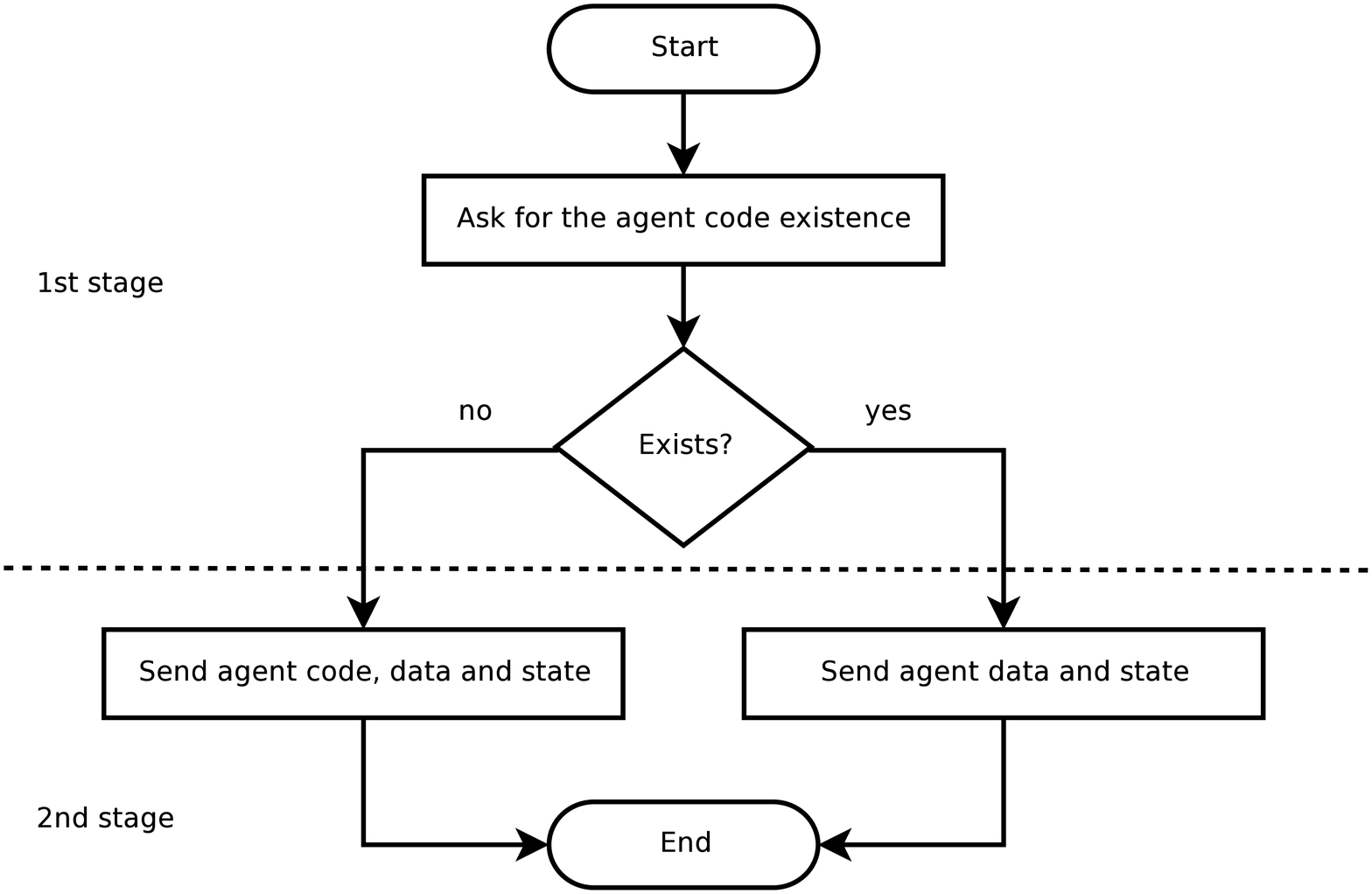}
  \caption{Push Transfer Diagram.}
  \label{fig:transfer-diagram}
\end{figure}

The protocol is divided in two stages (see
Figure~\ref{fig:transfer-diagram}). In the first stage a code unique
identifier (CID) value (generated by a cryptographic hash function) is
sent to ask whether the agent code transfer is needed. In the second
stage, the code, the agent data, and the agent state (when it is
needed) are sent. The ontology used in this protocol is called
``push-transfer-protocol-ontology-v1''.

The first part uses an IEEE-FIPA Proposal Interaction
Protocol~\cite{FIPA036} (see Figure~\ref{fig:fipa-proposal}). The
first message sent contains a \texttt{negotiate} predicate (see that
on Table~\ref{tab:negotiate-predicate}) with the code unique
identification (CID) value (see
Table~\ref{tab:push-transfer-ontology-negotiate}). Then a response
message shall be received accepting the proposal to send the code or
rejecting it. In case of error, the proposal is rejected and an error
is included as a message content.

\begin{table*}
  \centering
  \begin{tabular}{|p{60pt}|p{275pt}|}
    \hline
    \textbf{Predicate} & negotiate \\
    \hline
    \textbf{Ontology} & push-transfer-protocol-ontology-v1 \\
    \hline
    \textbf{Supported by} & amm \\
    \hline
    \textbf{Description} & Propose to negotiate whether send the agent code to the destination platform. \\
    \hline
    \textbf{Domain} & push-transfer-protocol-negotiate \\
    \hline
  \end{tabular}
  \caption{Push Transfer Protocol Ontology: Negotiate predicate.}
  \label{tab:negotiate-predicate}
\end{table*}

\begin{table*}
  \centering
  \begin{tabular}{|p{60pt}|p{150pt}|p{50pt}|p{75pt}|}
    \hline
    \multicolumn{1}{|l|}{\textbf{Frame}} &
    \multicolumn{3}{|l|}{push-transfer-protocol-negotiate} \\
    \hline
    \multicolumn{1}{|l|}{\textbf{Ontology}} &
    \multicolumn{3}{|l|}{push-transfer-protocol-ontology-v1} \\
    \hline
    \textbf{Parameter}&\textbf{Description}&\textbf{Presence}&\textbf{Type} \\
    \hline
    cid & Agent code unique identifier (CID) & Optional & String \\
    \hline
  \end{tabular}
  \caption{Push Transfer Protocol Ontology: negotiate transfer.}
  \label{tab:push-transfer-ontology-negotiate}
\end{table*}

The second part uses a simplified FIPA Request Interaction
Protocol~\cite{FIPA026} (see Figure~\ref{fig:fipa-request}). The first
message sent contains a \texttt{transfer} action (as seen in
Table~\ref{tab:transfer-action}) with the code, data and/or state of
the migrating agent (see Table~\ref{tab:push-transfer-ontology}). It
must be noted that the code, data and state are packed according to
the specific mobile agent system. For example, in a JADE system the
code is placed inside a JAR file, the data is on a byte array
resulting from the Java agent object serialisation, and the state is
not used. Once the agent is transferred, the sending platform expects
to receive an \texttt{inform-done} message, if the agent has been
correctly installed in the destination platform, or a
\texttt{failure}, otherwise.

\begin{table*}
  \centering
  \begin{tabular}{|p{60pt}|p{275pt}|}
    \hline
    \textbf{Function} & transfer \\
    \hline
    \textbf{Ontology} & push-transfer-protocol-ontology-v1 \\
    \hline
    \textbf{Supported by} & amm \\
    \hline
    \textbf{Description} & Request to send the agent code and instance to the destination platform. \\
    \hline
    \textbf{Domain} & push-transfer-protocol-transfer \\
    \hline
  \end{tabular}
  \caption{Push Transfer Protocol Ontology: Transfer action.}
  \label{tab:transfer-action}
\end{table*}

\begin{table*}
  \centering
  \begin{tabular}{|p{60pt}|p{150pt}|p{50pt}|p{75pt}|}
    \hline
    \multicolumn{1}{|l|}{\textbf{Frame}} &
    \multicolumn{3}{|l|}{push-transfer-protocol-transfer} \\
    \hline
    \multicolumn{1}{|l|}{\textbf{Ontology}} &
    \multicolumn{3}{|l|}{push-transfer-protocol-ontology-v1} \\
    \hline
    \textbf{Parameter}&\textbf{Description}&\textbf{Presence}&\textbf{Type} \\
    \hline
    cid & Agent code unique identifier (CID) & Optional & String \\
    \hline
    code & Agent code & Optional & Byte-Stream \\
    \hline
    data & Agent data & Mandatory & Byte-Stream \\
    \hline
    state & Agent state & Optional  & Byte-Stream \\
    \hline
  \end{tabular}
  \caption{Push Transfer Protocol Ontology: transfer agent.}
  \label{tab:push-transfer-ontology}
\end{table*}

\section{Conclusions}
\label{sec:conclusions}

Mobile agent systems require new open and interoperable mechanisms for
migration. In order to face this problem, we have presented a new open
architecture for the future standardisation of agent migration based
on several IEEE-FIPA agent specifications.

This architecture splits the migration process into three steps
(Pre-Transfer, Transfer, and Post-Transfer), with a flexible open set
of protocols in each step, plus two additional fixed steps (Agent
Registration and Agent Power Up), each one implemented by an already
defined protocol. The concrete protocols for the first three steps are
deliberately left unspecified so that the architecture can accommodate
a wide range of protocols and strategies.
The most important of these three steps is the Transfer, actually the
only one required among the three of them. As an example of Transfer
protocol we have also presented the Push Transfer Protocol.

To demonstrate the feasibility of our proposal, we have successfully
implemented and tested it in two different agent platforms: JADE and
AgentScape. In the case of the JADE agent platform, the new migration
architecture is the evolution of the former migration
service~\cite{acmnr0906}. This can be downloaded from the JIPMS
SourceForge project website (\texttt{http://jipms.sourceforge.net}) as
from the development release 1.98. The proposed migration has also
been implemented into the AgentScape platform, once it has been
successfully tested in JADE. This has involved the implementation of
several FIPA specifications because AgentScape is not a FIPA compliant
platform.  These implementations prove that the proposed architecture
is valid for completely different agent platforms.

Despite that these two platforms have the same migration architecture,
agents cannot move between them because the agent Application
Programming Interfaces are different and, therefore, incompatible. As
a future work, part of our research will be focused on migrating and
running agents from a specific platform technology to a different one.
The first step, that is, the definition of mobility interoperable
mechanisms for different platforms, has already been solved by this
proposal.

Furthermore, another future work will be the research on new
protocols, like new migration schemes (on demand migration, fragmented
migration, ...), authentication and authorisation mechanisms, agent
results transferring, etc., all of them im\-ple\-men\-ta\-ble in one
of the first three steps of the migration proposal.

\section*{Acknowledgments}
We want to thank J.R. Velasco and I.Marsa of the Universidad de
Alcal\'a de Henares for their suggestions in the architecture
proposed. This work has been partially funded through the Spanish
national project TSI2006-03481, and the support of the Catalan project
2005-FI-00752 and the European Social Fund (ESF).

\bibliographystyle{plain}
\bibliography{local}

\begin{thebibliography}{10}

\bibitem{acmnr0906}
J.~Ametller, J.~Cucurull, R.~Martí, G.~Navarro, and S.~Robles.
\newblock Enabling mobile agents interoperability through fipa standards.
\newblock In {\em Cooperative Information Agents X}, volume 4149 of {\em LNAI},
  pages 388--401. Springer Verlag, 2006.

\bibitem{br05}
P.~Braun and W.~Rossak.
\newblock {\em Mobile Agents}.
\newblock Morgan Kaufmann and dpunkt.verlag, 2005.

\bibitem{FIPA087}
FIPA.
\newblock Fipa agent management support for mobility specification, 2000.

\bibitem{FIPA061}
FIPA.
\newblock Fipa acl message structure specification, 2002.

\bibitem{FIPA036}
FIPA.
\newblock Fipa propose interaction protocol specification, 2002.

\bibitem{FIPA026}
FIPA.
\newblock Fipa request interaction protocol specification, 2002.

\bibitem{FIPA023}
FIPA.
\newblock Fipa agent management specification, 2004.

\bibitem{586291}
R.~S. Gray, G.~Cybenko, D.~Kotz, R.~A. Peterson, and D.~Rus.
\newblock D'agents: applications and performance of a mobile-agent system.
\newblock {\em Softw. Pract. Exper.}, 32(6):543--573, 2002.

\bibitem{JADE}
{JADE, Java Agent DEvelopment Framework}.
\newblock \texttt{http://jade.cselt.it}, 2007.

\bibitem{JK00}
W.~Jansen and T.~Karygiannis.
\newblock Nist special publication 800-19 - mobile agent security, 2000.

\bibitem{lo98}
D.~B. Lange and M.~Oshima.
\newblock Mobile agents with java: The aglet api.
\newblock {\em World Wide Web}, 1(3):111--121, 1998.

\bibitem{MASIF}
{OMG Mobile Agent Systems Interoperability Facilities Specification (MASIF),
  OMG TC Document ORBOS/97-10-05 }.

\bibitem{Overeinder.ea:04}
B.~J. Overeinder and F.~M.~T. Brazier.
\newblock Scalable middleware environment for agent-based internet
  applications.
\newblock In {\em Proceedings of the Workshop on State-of-the-Art in Scientific
  Computing}, volume 3732 of {\em LNCS}, pages 675--679. Springer, 2004.

\bibitem{btk05}
D.~Trinh P.~Braun and R.~Kowalczyk.
\newblock Integrating a new mobility service into the jade agent toolkit.
\newblock In {\em Mobility Aware Technologies and Applications}, volume 3744 of
  {\em LNCS}, pages 354--363. Springer, 2005.

\bibitem{732412}
H.~Peine and T.~Stolpmann.
\newblock The architecture of the ara platform for mobile agents.
\newblock In {\em Proceedings of the First International Workshop on Mobile
  Agents}, pages 50--61. Springer, 1997.

\bibitem{Pinsdorf:2002a}
U.~Pinsdorf and V.~Roth.
\newblock {Mobile Agent Interoperability Patterns and Practice}.
\newblock In {\em Proceedings of Ninth IEEE International Conference and
  Workshop on the Engineering of Computer-Based Systems}, pages 238--244. IEEE
  Computer Society Press, 2002.

\bibitem{roth01}
V.~Roth.
\newblock On the robustness of some cryptographic protocols for mobile agent
  protection.
\newblock In {\em Mobile Agents: 5th International Conference}, volume 2240 of
  {\em LNCS}, 2001.

\bibitem{VRCCNM525}
P.~Vieira-Marques, S.~Robles, J.~Cucurull, R.~Cruz-Correia, G.~Navarro, and
  R.~Martí.
\newblock Secure integration of distributed medical data using mobile agents.
\newblock {\em IEEE Intelligent Systems}, 21(6), 2006.

\bibitem{white96}
J.~E. White.
\newblock Telescript technology: Mobile agents.
\newblock In Jeffrey Bradshaw, editor, {\em Software Agents}. AAAI Press/MIT
  Press, 1996.

\end{thebibliography}

\end{document}